\documentclass[a4paper]{article}

\usepackage{amsfonts}
\usepackage{amssymb}
\usepackage{mathrsfs}
\usepackage{float}
\usepackage{amsmath}
\usepackage{amsfonts,amssymb}
\usepackage{graphicx,color}
\usepackage{mathrsfs}

\newtheorem{Theorem}{Theorem}[section]

\newtheorem{Lemma}{Lemma}[section]

\newtheorem{Corollary}{Corollary}[section]
\newtheorem{Remark}{Remark}[section]

\newtheorem{Proposition}{Proposition}[section]

\catcode`\@=11 \@addtoreset{equation}{section}

\catcode`\@=12

\date{ }
\title{The motion of relativistic strings in curved space-times}
\author{Chun-Lei He\footnote{Department of Mathematics, Anhui Normal
University, Wuhu 241000, China;}  $\quad$ and $\quad$ De-Xing
Kong\footnote{Department of Mathematics, Zhejiang University,
Hangzhou 310027, China;} \footnote{ Corresponding author:
kong@cms.zju.edu.cn.}
\\}

\begin{document}
\maketitle
\begin{abstract}
This paper concerns the motion of a relativistic string in a curved
space-time. As a general framework, we first analyze relativistic
string equations, i.e., the basic equations for the motion of a
one-dimensional extended object in a curved enveloping space-time
$(\mathscr{N}, \tilde g)$, which is a general Lorentzian manifold,
and then investigate some interesting properties enjoyed by these
equations. Based on this, under suitable assumptions we prove the
global existence of smooth solutions of the Cauchy problem for
relativistic string equations in the curved space-time
$(\mathscr{N}, \tilde g)$. In particular, we consider the motion of
a relativistic string in the Ori's space-time, and give a sufficient
and necessary condition guaranteeing the global existence of smooth
solutions of the Cauchy problem for relativistic string equations in
the Ori's space-time.\vskip 6mm


\vskip 3mm


\end{abstract}

\newpage
\baselineskip=7mm

\section{Introduction}
This paper concerns the nonlinear dynamics of a relativistic string
moving in a curved space-time. It is well known that, in particle
physics, the string model is frequently used to study the structure
of hardrons. In fact, a free string is a one-dimensional physical
object whose motion is represented by a time-like surface. On the
other hand, in mathematics, extremal surfaces in a physical
space-time include the following four types: space-like, time-like,
light-like or mixed types. For the case of space-like extremal
(minimal or maximal) surfaces in the Minkowski space-time, we refer
to the classical papers by Calabi \cite{c} and by Cheng and Yau
\cite{cy}. The case of time-like extremal surfaces in the Minkowski
space-time has been investigated by several authors (e.g., \cite{ba}
and \cite{m}). Barbashov, Nesterenko and Chervyakov in \cite{ba}
study nonlinear partial differential equations describing extremal
surfaces in the Minkowski space-time and provide examples with exact
solutions. Milnor \cite{m} generates examples that display
considerable variety in the shape of entire time-like extremal
surfaces in the 3-dimensional Minkowski space-time
$\mathbb{R}^{1+2}$ and shows that such surfaces need not be planar.
For the case of extremal surfaces of mixed type, Gu investigates the
extremal surfaces of mixed type in the $n$-dimensional Minkowski
space-time (cf. \cite{g3}) and constructs many complete extremal
surfaces of mixed type in the 3-dimensional Minkowski space-time
(cf. \cite{g4}). Recently, Kong et al re-study the equations for
time-like extremal surfaces in the Minkowski space-time
$\mathbb{R}^{1+n}$, which corresponds to the motion of an open
string in $\mathbb{R}^{1+n}$ (see \cite{k}-\cite{ksz}). For the
multidimensional version, Hoppe et al derive the equation for a
classical relativistic membrane moving in the Minkowski space-time
$\mathbb{R}^{1+3}$, which is a nonlinear wave equation corresponding
to the extremal hypersurface equation in $\mathbb{R}^{1+3}$, and
give some special classical solutions (cf. \cite{bh}, \cite{h}). The
Cauchy problem with small initial data for the time-like extremal
surface equation in the Minkowski space-time has been studied
successfully by Lindblad \cite{lin} and, by Chae and Huh \cite{CH}
in a more general framework. Using the null forms in Christodoulou
and Klainerman's style (cf. \cite{ch} and \cite{kla}), they prove
the global existence of smooth solutions for sufficiently small
initial data with compact support.

In \cite{kzz}, Kong et al investigate the dynamics of relativistic
(in particular, closed) strings moving in the multidimensional
Minkowski space-time $\mathbb{R}^{1+n}\;(n\ge 2)$. They first derive
a system with $n$ nonlinear wave equations of Born-Infeld type which
governs the motion of the string. This system can also be used to
describe the extremal surfaces in $\mathbb{R}^{1+n}$. Then they show
that this system enjoys some interesting geometric properties. Based
on this, they give a sufficient and necessary condition guaranteeing
the global existence of extremal surfaces without space-like point
for given initial data. This result corresponds to the global
propagation of nonlinear waves for the system describing the motion
of the string in $\mathbb{R}^{1+n}$. Moreover, a great deal of
numerical analysis are investigated, and the numerical results show
that, in phase space, various topological singularities develop in
finite time in the motion of the string. More recently, Kong and
Zhang furthermore study the motion of relativistic strings in the
Minkowski space $\mathbb{R}^{1+n}$ (see \cite{kz}). Surprisingly,
they obtain a general solution formula for this complicated system
of nonlinear wave equations. Based on this solution formula, they
successfully prove that the motion of closed strings is always
time-periodic. Moreover, they further extend the solution formula to
finite relativistic strings.

However, in a curved space-time there are only few results to obtain
(see \cite{g1} and Sections 24 and 32 in \cite{Barbashov}). Gu
\cite{g1} shows that the motion of a string can be determined by
constructing a certain wave map from the Minkowski plane to the
enveloping space-time which is a given Lorentzian manifold. Later,
Gu investigates the Cauchy problem for wave maps from
$\mathbb{R}^{1+1}$ to $\mathbb{S}^{1+1}$ and proves a theorem on the
existence of global smooth solutions (see \cite{g10}). Recently, in
\cite{hek} we consider the motion of relativistic strings in the
Schwarzschild space-time, and under suitable assumptions we prove a
global existence theorem on smooth solutions of the Cauchy problem
for the equations for the motion of relativistic strings with {\it
small} arc length.

In this paper we consider the nonlinear dynamics of a relativistic
string moving in a curved space-time. As a general framework, we
first analyze relativistic string equations, i.e., the basic
equations for the motion of a one-dimensional extended object in a
curved enveloping space-time $(\mathscr{N}, \tilde g)$ which stands
for a general Lorentzian manifold, and then investigate some
interesting properties enjoyed by these equations. Based on this,
under suitable assumptions we prove the global existence of smooth
solutions of the Cauchy problem for relativistic string equations in
the curved space-time $(\mathscr{N}, \tilde g)$. Since the Ori's
space-time has recently received much attention mainly due to the
fact that it is a time-machine solution with a compact vacuum core
of the Einstein's field equations (see \cite{ori}), we particularly
consider the motion of a relativistic string in the Ori's
space-time, and give a sufficient and necessary condition
guaranteeing the global existence of smooth solutions of the Cauchy
problem for relativistic string equations in the Ori's space-time.

The paper is organized as follows. In Section 2, we investigate the
nonlinear dynamics of a relativistic string moving in a general
curved space-time. Section 3 is devoted to the study on the motion
of a relativistic string in the Ori's space-time. The conclusion and
discussion are given in Section 4.

\section{The motion of relativistic strings in general curved space-times}
In this section, we investigate the motion of a one-dimensional
extended object in the enveloping space-time $(\mathscr{N}, \tilde
g)$, which stands for a given general Lorentzian manifold.

Since the world sheet of the one-dimensional extended object
corresponds to a two-dimensional extremal sub-manifold, denoted by
$\mathscr{M}$, we may choose the local coordinates $(\zeta^0,
\zeta^1)$ in $\mathscr{M}$. For simplicity, we also denote
$\zeta^0=t, \zeta^1=\theta$. Let the position vector in the
space-time $(\mathscr{N}, \tilde g)$ be
\begin{equation}\label{2.1}X(t,\theta)=(x^0(t,\theta),x^1(t,\theta),\cdots,
x^n(t,\theta)).\end{equation} Denote
\begin{equation}\label{2.2}x^A_{\mu}=\dfrac{\partial x^A}{\partial \zeta^{\mu}}\quad
{\rm{and}}\quad x^A_{\mu\nu}=\dfrac{\partial^2 x^A}{\partial
\zeta^\mu
\partial \zeta^{\nu}}\quad (A=0,1,\cdots,n;\;\mu,\,\nu=0,1).\end{equation} Then the induced metric of the sub-manifold
$\mathscr{M}$ can be written as $g=(g_{\mu\nu})$, where
\begin{equation}\label{2.3}g_{\mu\nu}=\tilde g_{AB}x^A_{\mu}x^B_{\nu}\quad (A,B=0,1,\cdots,n;\;\mu,\nu=0,1).\end{equation}
As a result, the corresponding Euler-Lagrange equations for the
one-dimensional extended object moving in the space-time
$(\mathscr{N}, \tilde g)$ read
\begin{equation}\label{2.4}g^{\mu\nu}\left(x^C_{\mu\nu}+\tilde\Gamma_{AB}^{C}x^A_{\mu}x^B_{\nu}-\Gamma_{\mu\nu}
^{\rho}x^C_{\rho}\right)=0\quad (C=0,1,\cdots,n),\end{equation}
where $g^{-1}\triangleq(g^{\mu\nu})$ is the inverse of the metric
$g$, $\tilde\Gamma_{AB}^{C}$ and $\Gamma_{\mu\nu}^{\rho}$ stand for
the connections of the metric $\tilde g$ and the induced metric $g$,
respectively.

Since we are only interested in the physical motion, we
may assume that the sub-manifold $\mathscr{M}$ is $C^2$ and {\it
time-like}, i.e.,
\begin{equation}\label{2.5} \Delta\triangleq\det g<0.\end{equation}
This implies that the world sheet of the extended object is
time-like, and then the motion satisfies the causality, or say, the
motion is physical.

Under the assumption (\ref{2.5}), the global solution to (\ref{2.4})
is diffeomorphic to the global solution of the following equations
provided with the same initial data (see \cite{ac} and \cite{hh})
\begin{equation}\label{2.6}E_C\triangleq g^{\mu\nu}\left(x^C_{\mu\nu}+\tilde\Gamma_{AB}^{C}
x^A_{\mu}x^B_{\nu}\right)=0 \quad (C=0,1,\cdots,n).\end{equation} So
in this paper it suffices to investigate the global existence of
smooth solutions of the system (\ref{2.6}) instead of (\ref{2.4}).

\begin{Remark}
The mapping (\ref{2.1}) described by the system (\ref{2.6}) is essentially a wave map from
the Minkowski space $\mathbb{R}^{1+1}$ to the Lorentzian manifold
$(\mathscr{N}, \tilde g)$. Gu \cite{g2} proved that the solution of
the Cauchy problem for the harmonic map
$\phi:\;\mathbb{R}^{1+1}\rightarrow \mathcal{M}$ exists globally,
where $\mathcal{M}$ is a Riemannian manifold. According to the
authors' knowledge, there exists only a few results on the wave map
from $\mathbb{R}^{1+1}$ to a general Lorentzian
manifold.\end{Remark}

Notice that the system (\ref{2.6}) can be written in the following
form
\begin{equation}\label{2.7}g_{11}x^C_{tt}-2g_{01}x^C_{t\theta}+g_{00}x^C_{\theta\theta}+g_{11}
\tilde{\Gamma}^C_{AB}x_t^Ax_t^B
-2g_{01}\tilde{\Gamma}^C_{AB}x_t^Ax_{\theta}^B+g_{00}\tilde{\Gamma}^C_{AB}x_{\theta}^Ax_{\theta}^B=0.
\end{equation}
Introduce
\begin{equation}\label{2.8}u=X,\;\;v=X_t,\;\;w=X_{\theta}
\end{equation}
and denote
\begin{equation}\label{2.9}U\triangleq (u,v,w)^T.
\end{equation}
Then the system (\ref{2.7}) can be equivalently rewritten as
\begin{equation}\label{2.10}U_t+AU_{\theta}+B=0,
\end{equation}
where
\begin{equation}\label{2.11}A=\left(\begin{array}{llllllrrrrrr}&0&0&0\\&0&-\dfrac{2g_{01}}{g_{11}}
I_{n+1}&\dfrac{g_{00}}{g_{11}}I_{n+1}\\
&0&-I_{n+1}&0\end{array}\right)_{3(n+1)\times 3(n+1)}\end{equation} and
\begin{equation}\label{2.12}B=(-v^T, \bar{B}^T,0)^T_{\;3(n+1)\times 1},\end{equation}
in which $\bar{B}=(\bar{B}^0, \bar{B}^1, \cdots, \bar{B}^{n})^{T}$ and
\begin{equation}\label{2.13}\begin{aligned}\bar{B}^C&=\tilde{\Gamma}^C_{AB}x_t^Ax_t^B-\frac{2g_{01}}
{g_{11}}\tilde{\Gamma}^C_{AB}x_t^Ax_{\theta}^B+
\frac{g_{00}}{g_{11}}\tilde{\Gamma}^C_{AB}x_{\theta}^Ax_{\theta}^B\\
&=\tilde{\Gamma}^C_{AB}v^Av^B-\frac{2g_{01}}{g_{11}}\tilde{\Gamma}^C_{AB}v^Aw^B+
\frac{g_{00}}{g_{11}}\tilde{\Gamma}^C_{AB}w^Aw^B.
\end{aligned}\end{equation}
By a direct calculation, the eigenvalues of the matrix $A$ read
\begin{equation}\label{2.14}\left\{\begin{aligned}&\lambda_1=\cdots=\lambda_{n+1}\triangleq
\lambda_0=0,\\
&\lambda_{n+2}=\lambda_{n+3}=\cdots=\lambda_{2n+2}\triangleq\lambda_-=\frac{-g_{01}-
\sqrt{g_{01}^2-g_{00}g_{11}}}{g_{11}},\\
&\lambda_{2n+3}=\lambda_{2n+4}=\cdots=\lambda_{3n+3}\triangleq\lambda_+=
\frac{-g_{01}+\sqrt{g_{01}^2-g_{00}g_{11}}}{g_{11}}.\end{aligned}\right.\end{equation} The right
eigenvector corresponding to $\lambda_i \; (i=1,2,\cdots,3n+3)$ can be
chosen as
\begin{equation}\label{2.15}\left\{\begin{aligned}&r_i=(e_i,0,0)^T\quad (i=1,\cdots,n+1),\\&
r_i=(0,-\lambda_{-}e_{i-(n+1)},e_{i-(n+1)})^T\quad
(i=n+2,\cdots,2n+2),\\&
r_i=(0,-\lambda_{+}e_{i-(2n+2)},e_{i-(2n+2)})^T\quad
(i=2n+3,\cdots,3n+3),\end{aligned}\right.\end{equation}
where$$e_{i}=(0,\cdots,0,\overset{(i)}{1},0,\cdots,0)
\quad(i=1,\cdots,n+1).$$ While, the left eigenvector corresponding
to $\lambda_i\; (i=1,2,\cdots,3n+3)$ can be taken as
\begin{equation}\label{2.16}\left\{\begin{aligned}&l_i=(e_i,0,0)\quad (i=1,\cdots,n+1),\\&
l_i=(0,e_{i-(n+1)},\lambda_{+}e_{i-(n+1)})\quad (i=n+2,\cdots,
2n+2),\\& l_i=(0,e_{i-(2n+2)},\lambda_{-}e_{i-(2n+2)})\quad
(i=2n+3,\cdots,3n+3).\end{aligned}\right.\end{equation}

\begin{Proposition}Under the assumption (\ref{2.5}), the system (\ref{2.10}) is a non-strictly
hyperbolic system with $3(n+1)$ eigenvalues (see (\ref{2.14})), and
the corresponding right (resp. left) eigenvectors can be chosen as
(\ref{2.15}) (resp. (\ref{2.16})).\end{Proposition}

\begin{Proposition}Under the assumption (\ref{2.5}), the system (\ref{2.10}) is linearly degenerate in
the sense of Lax (see \cite{lax}).\end{Proposition}

\noindent{\bf Proof.} Obviously, it holds that
$$\nabla\lambda_0\cdot r_i=0\quad (i=1,2,\cdots,n+1).$$
We next calculate the invariants $\nabla\lambda_-\cdot
r_i\;\,(i=n+2,\cdots,2n+2)$ and $\nabla\lambda_+\cdot
r_i\;\,(i=2n+3,\cdots,3n+3).$

In fact, for every $C\in\{0,1,2,\cdots,n\}$, by a direct calculation
we obtain
\begin{equation}\label{2.17}\frac{\partial \lambda_-}{\partial v^C}=\tilde{g}_{CB}w^B\frac{\lambda_-}
{\sqrt{g_{01}^2-g_{00}g_{11}}}+
\tilde{g}_{CB}v^B\frac{1}{\sqrt{g_{01}^2-g_{00}g_{11}}}\end{equation}
and
\begin{equation}\label{2.18}\frac{\partial \lambda_-}{\partial w^C}=\tilde{g}_{CB}w^B
\frac{\lambda_-^2}{\sqrt{g_{01}^2-g_{00}g_{11}}}+
\tilde{g}_{CB}v^B\frac{\lambda_-}{\sqrt{g_{01}^2-g_{00}g_{11}}}.\end{equation}
Then we have
\begin{equation}\label{2.19}\nabla\lambda_-\cdot r_{n+2+C}=-\lambda_-\frac{\partial\lambda_-}
{\partial v^C}+\frac{\partial\lambda_-}{\partial w^C}=0\quad
(C=0,1,2,\cdots,n).
\end{equation}
Similarly, we can prove
\begin{equation}\label{2.20}\nabla\lambda_+\cdot r_{2n+3+C}=-\lambda_+\frac{\partial\lambda_+}
{\partial v^C}+\frac{\partial\lambda_+}{\partial w^C}=0\quad
(C=0,1,2,\cdots,n).
\end{equation}
Thus, the proof is completed. $\qquad\qquad\square$

\begin{Theorem}Under the assumption (\ref{2.5}), $\lambda_{-}$ (resp. $\lambda_{+}$) is a Riemann invariant
corresponding to $\lambda_{+}$ (resp. $\lambda_{-}$). Moreover,
these two Riemann invariants satisfy
\begin{equation}\label{2.21}\dfrac{\partial\lambda_{-}}{\partial t}+\lambda_{+}\dfrac{\partial\lambda_{-}}
{\partial\theta}=0,\quad \dfrac{\partial\lambda_{+}}{\partial
t}+\lambda_{-}\dfrac
{\partial\lambda_{+}}{\partial\theta}=0.\end{equation}\end{Theorem}

\noindent{\bf Proof.} Multiplying (\ref{2.10}) by the left
eigenvectors given by (\ref{2.16}) leads to
\begin{equation}\label{2.22}\left\{\begin{aligned}v^C_t+\lambda_-v^C_{\theta}+\lambda_+(w_t^C+\lambda_-w_{\theta}^C)+
\bar{B}^C=0,\\
v^C_t+\lambda_+v^C_{\theta}+\lambda_-(w_t^C+\lambda_+w_{\theta}^C)+\bar{B}^C=0.
\end{aligned}\right.\end{equation}
Noting (\ref{2.19}) and using (\ref{2.22}), we have
\begin{equation}\label{2.23}\begin{aligned}\dfrac{\partial\lambda_{-}}{\partial t}+
\lambda_{+}\dfrac{\partial\lambda_{-}}
{\partial\theta}&=\frac{\partial\lambda_-}{\partial u^C}v^C+\lambda_+\frac{\partial\lambda_-}{\partial u^C}w^C+
\frac{\partial\lambda_-}{\partial v^C}(v^C_t+\lambda_+v^C_{\theta})+\frac{\partial\lambda_-}{\partial w^C}(w^C_t+
\lambda_+w^C_{\theta})\\
&=\frac{\partial\lambda_-}{\partial u^C}v^C+\lambda_+\frac{\partial\lambda_-}{\partial u^C}w^C+
\frac{\partial\lambda_-}{\partial v^C}(v^C_t+\lambda_+v^C_{\theta})+
\frac{\partial\lambda_-}{\partial v^C}\lambda_-(w^C_t+\lambda_+w^C_{\theta})\\
&=\frac{\partial\lambda_-}{\partial u^C}(v^C+\lambda_+w^C)-\frac{\partial\lambda_-}{\partial v^C}\bar{B}^C
\end{aligned}\end{equation}
By a direct calculation,
\begin{equation}\label{2.24}\frac{\partial \lambda_-}{\partial
u^C}=\frac{1}{2\sqrt{g_{01}^2-g_{00}g_{11}}} \frac{\partial
\tilde{g}_{AB}}{\partial
u^C}(v^A+\lambda_-w^A)(v^B+\lambda_-w^B).\end{equation} On the other
hand, notice that (\ref{2.13}) can be rewritten as
\begin{equation}\label{n1}\bar{B}^C=\tilde{\Gamma}^C_{AB}(v^A+\lambda_-w^A)(v^B+\lambda_+w^B).\end{equation}
Then, substituting (\ref{2.17}), (\ref{2.24}) and (\ref{n1}) into (\ref{2.23}) yields
$$\dfrac{\partial\lambda_{-}}{\partial t}+\lambda_{+}\dfrac{\partial\lambda_{-}}
{\partial\theta}=0.$$
 Similarly, we can prove
 $$\dfrac{\partial\lambda_{+}}{\partial
t}+\lambda_{-}\dfrac {\partial\lambda_{+}}{\partial\theta}=0.$$
Thus, the proof is completed. $\qquad\qquad\square$

\begin{Remark} The system (\ref{2.21}) is a $2\times 2$ quasilinear hyperbolic system with
linearly degenerate characteristic fields, it plays an important
role in our argument.\end{Remark}

Introduce
\begin{equation}\label{2.25}
p^C=v^C+\lambda_-w^C,\quad q^C=v^C+\lambda_+w^C\quad
(C=0,1,\cdots,n).
\end{equation}
Then the system (\ref{2.10}) can be equivalently rewritten as
\begin{equation}\label{2.26}\left\{\begin{aligned}\frac{\partial p^C}{\partial t}+\lambda_+\frac{\partial p^C}
{\partial \theta}=-\tilde{\Gamma}
^C_{AB}(u)p^Aq^B,\\
\frac{\partial q^C}{\partial t}+\lambda_-\frac{\partial
q^C}{\partial \theta}=-\tilde{\Gamma}
^C_{AB}(u)p^Aq^B\end{aligned}\qquad (C=0,1,\cdots,n). \right.
\end{equation}
\begin{Remark} Noting (\ref{2.21}) and (\ref{2.26}), we observe
that, once one can solve $\lambda_{\pm}$ from the system
(\ref{2.21}), then (\ref{2.26}) becomes a semilinear hyperbolic
system of first order.
\end{Remark}

Moreover, by calculations we have
$$\begin{aligned}\tilde{g}_{AB}p^Ap^B=& \;\tilde{g}_{AB}(v^A+\lambda_-w^A)(v^B+\lambda_-w^B)\\
=& \;\tilde{g}_{AB}v^Av^B+\lambda_-\tilde{g}_{AB}w^Av^B+
\lambda_-\tilde{g}_{AB}v^Aw^B+\lambda_-^2\tilde{g}_{AB}w^Aw^B\\
=& \;g_{00}+2\lambda_-g_{01}+\lambda_-^2g_{11}\equiv0\end{aligned}$$
and
$$\tilde{g}_{AB}q^Aq^B\equiv 0.$$
Thus we have proved
\begin{Proposition} $p$ and $q$ are two null vectors, i.e., it holds
that
\begin{equation}\label{2.27}\tilde{g}_{AB}p^Ap^B\equiv0 \quad and \quad \tilde{g}_{AB}q^Aq^B\equiv0.\end{equation}
\end{Proposition}

At the end of this section, we consider the global existence of
smooth solutions of the relativistic string equations in a general
curved space-time.

Consider the Cauchy problem for the system (\ref{2.6}) (or
equivalently, (\ref{2.7})) with the initial data
\begin{equation}\label{2.28}x^C(0,\theta)=\varphi^C(\theta),\quad x_t^C(0,\theta)=\psi^C(\theta)\quad (C=0,1,\cdots,n),
\end{equation}
where $\varphi^C(\theta)$ are $C^2$-smooth functions with bounded
$C^2$-norm, while $\psi^C(\theta)$ are $C^1$-smooth functions with
bounded $C^1$-norm. In physics,
$\varphi(\theta)=(\varphi^0(\theta),\varphi^1(\theta),\cdots,\varphi^n(\theta))$
and $\psi(\theta)=(\psi^0(\theta),
\psi^1(\theta),\cdots,\psi^n(\theta))$ stand for the initial
position and the initial velocity of the string under consideration,
respectively.

Introduce
\begin{equation}\label{2.29}
\Lambda_{\pm}(\theta)=\frac{-g_{01}[\varphi,\psi](\theta)\pm
\sqrt{\left(g_{01}[\varphi,\psi](\theta)\right)^2-g_{00}[\varphi,\psi](\theta)g_{11}[\varphi,\psi](\theta)}
}{g_{11}[\varphi,\psi](\theta)}
\end{equation}
and
\begin{equation}\label{2.30}\mathcal{L}(\theta)=g_{00}[\varphi,\psi](\theta)g_{11}[\varphi,\psi](\theta)-
\left(g_{01}[\varphi,\psi](\theta)\right)^2,
\end{equation}
where
\begin{equation}\label{2.31}
g_{00}[\varphi,\psi](\theta)=\tilde{g}_{AB}(\varphi)\psi^A\psi^B,\quad
 g_{01}[\varphi,\psi](\theta)=\tilde{g}_{AB}(\varphi)\psi^A\varphi_{\theta}^B,\quad
 g_{11}[\varphi,\psi](\theta)=\tilde{g}_{AB}(\varphi)\varphi^A_{\theta}\varphi^B_{\theta}.\end{equation}
In fact, in physics $\Lambda_{\pm}(\theta)$ stand for the
characteristic propagation speeds of the point $\theta$ at the
initial time, and $\mathcal{L}(\theta)$ denotes the Lagrangian
energy density.

For given $\varphi$ and $\psi$, we consider the Cauchy problem for
the system (\ref{2.21}) with the initial data
\begin{equation}\label{2.32}t=0:\quad
\lambda_{\pm}=\Lambda_{\pm}(\theta),
\end{equation}
where $\Lambda_{\pm}(\theta)$ are defined by (\ref{2.29}). Since we
only consider the physical motion, it is natural to assume that
\begin{equation}\label{2.33}\Lambda_-(\theta)<\Lambda_+(\theta),\quad\forall \;\theta\in\mathbb{R}.
\end{equation}
The condition (\ref{2.33}) is equivalent to the fact that the
assumption (\ref{2.5}) is satisfied at the initial time, i.e., the
motion is physical at the time $t=0$.

Under the assumption (\ref{2.33}), by Kong and Tsuji \cite{kt}, the
Cauchy problem (\ref{2.21}), (\ref{2.32}) has a unique global $C^1$
solution $\lambda_{\pm}=\lambda_{\pm}(t,\theta)$ defined on
$\mathbb{R}^+\times\mathbb{R}$ if and only if, for every fixed
$\theta_2\in\mathbb{R}$, it holds that
\begin{equation}\label{2.34}\Lambda_-(\theta_1)<\Lambda_+(\theta_2),\quad\forall \;\theta_1<\theta_2.
\end{equation}
In fact, the condition (\ref{2.34}) guarantees that the motion is
always physical for all time $t\in \mathbb{R}^+$.

By the same method as in He and Kong \cite{hek}, we can prove the
following theorem.
\begin{Theorem}Suppose that $\tilde{g}$ is a Lorentzian metric, $\varphi(\theta)$ is a $C^2$-smooth vector-valued
function with bounded $C^2$-norm and $\psi(\theta)$ is a
$C^1$-smooth vector-valued function with bounded $C^1$-norm. Suppose
furthermore that the assumptions (\ref{2.33}) and (\ref{2.34}) are
satisfied. Then there exists a positive constant $\varepsilon$ such
that the Cauchy problem (\ref{2.6}), (\ref{2.28}) admits a unique
global $C^2$-smooth solution $x^C=x^C(t,\theta)$ for all
$t\in\mathbb{R}^+$, provided that
\begin{equation}\label{2.35}
\int^\infty_{-\infty}\left|\frac{d\varphi^C(\theta)}{d\theta}\right|d\theta\leq
\varepsilon \quad and\quad
\int^\infty_{-\infty}\left|\psi^C(\theta)\right|d\theta\leq
\varepsilon.
\end{equation}
\end{Theorem}
\begin{Remark} In Theorem 2.2, the constant $\varepsilon$ only depends on the $C^2$-norm
of $\varphi$ and the $C^1$-norm of $\psi$.
 The first inequality in (\ref{2.35})  implies that the $BV$-norm of
$\varphi^C(\theta)$ is small, that is, the arc length of the initial
string is small; while the second inequality in (\ref{2.35}) implies
that the $L^1$-norm of the initial velocity is small. The physical
meaning of Theorem 2.2 is as follows: for a string with small arc
length, the smooth motion exists globally (or say, no singularity
appears in the whole motion process), provided that the $L^1$-norm
of the initial velocity is small. In geometry, Theorem 2.2
essentially gives a global existence result on smooth solutions of a
wave map from the Minkowski space-time $\mathbb{R}^{1+1}$ to a
general curved space-time (cf. \cite{g1}).
\end{Remark}

\section{The motion of relativistic strings in Ori's space-time}
Since the Ori's space-time has recently received much attention
mainly due to the fact that it is a time-machine solution with a
compact vacuum core of the Einstein's field equations, in this
section we mainly investigate the motion of a relativistic string in
the Ori's space-time, and give a sufficient and necessary condition
guaranteeing the global existence of smooth solutions of the Cauchy
problem for relativistic string equations in the Ori's space-time.

As discussed in Section 2, under the assumption (\ref{2.33}), the
Cauchy problem (\ref{2.21}), (\ref{2.32}) has a unique global $C^1$
solution $\lambda_{\pm}=\lambda_{\pm}(t,\theta)$ on
$\mathbb{R}^+\times\mathbb{R}$ if and only if, for every fixed
$\theta_2\in\mathbb{R}$, (\ref{2.34}) holds. Moreover, on the
existence domain of the solution, it always holds that
\begin{equation}\label{3.0}
\lambda_{-}(t,\theta)<\lambda_{+}(t,\theta),\quad\forall\;
(t,\theta)\in \mathbb{R}^+\times \mathbb{R}.\end{equation} See Kong
and Tsuji \cite{kt}. Therefore, in what follows, we assume that
(\ref{2.33}) and (\ref{2.34}) are always satisfied.

On the other hand, it follows from Serre \cite{serre} that the
solution $(\lambda_-,\lambda_+)$ of the Cauchy problem (\ref{2.21}),
(\ref{2.32}) satisfies the following identity
\begin{equation}\label{3.1}
\partial_t\left(\frac{2}
{\lambda_+-\lambda_-}\right)
+\partial_{\theta}\left(\frac{\lambda_++\lambda_-}
{\lambda_+-\lambda_-}\right) =0.\end{equation}
This allows us to
introduce the following transformation of the variables
\begin{equation}\label{3.2}
(t,\theta) \longrightarrow (t,\vartheta),\end{equation}
where
$\vartheta=\vartheta(t,\theta)$ is given by
\begin{equation}\label{3.4}\left\{\begin{array}{l}{\displaystyle
d\vartheta=\frac{2}
{\lambda_+(t,\theta)-\lambda_-(t,\theta)}d\theta-
\frac{\lambda_+(t,\theta)+\lambda_-(t,\theta)}
{\lambda_+(t,\theta)-\lambda_-(t,\theta)}dt,\quad\forall\;
(t,\theta)\in \mathbb{R}^+\times \mathbb{R}} \vspace{3mm}\\
{\displaystyle\vartheta(0,\theta)=\Theta_0(\theta)\triangleq
\int^{\theta}_0\frac{2}{\Lambda_{+}(\zeta)-\Lambda_{-}(\zeta)}d\zeta,
\quad\forall\; \theta\in \mathbb{R}.}
\end{array}\right.\end{equation}

The following lemma comes from He and Kong \cite{hek}.
\begin{Lemma}Under the assumptions (\ref{2.33}) and (\ref{2.34}), the mapping defined by (\ref{3.2})-(\ref{3.4})
is globally diffeomorphic; moreover, it holds that
\begin{equation}\label{3.5}
\frac{\partial}{\partial t}+\lambda_+\frac{\partial}{\partial
\theta}=\frac{\partial}{\partial t}+\frac{\partial}{\partial
\vartheta},\quad \frac{\partial}{\partial
t}+\lambda_-\frac{\partial}{\partial
\theta}=\frac{\partial}{\partial t}-\frac{\partial}{\partial
\vartheta}.\end{equation}
\end{Lemma}

\begin{Remark}
The mapping defined by (\ref{3.2})-(\ref{3.4}) is somewhat similar
to the transformation between the Euler version and Lagrange version
for one-dimensional gas dynamics.
\end{Remark}

By Lemma 3.1, under the coordinates $(t,\vartheta)$, the system
(\ref{2.26}) can be equivalently rewritten as
 \begin{equation}\label{3.6}\left\{\begin{array}{l}\dfrac{\partial \bar{p}^C}{\partial t}+
 \dfrac{\partial \bar{p}^C}{\partial \vartheta}=
-\tilde{\Gamma}^C_{AB}(\bar{u})\bar{p}^A\bar{q}^B,\vspace{3mm}\\
\dfrac{\partial \bar{q}^C}{\partial t}-\dfrac{\partial
\bar{q}^C}{\partial \vartheta}=
-\tilde{\Gamma}^C_{AB}(\bar{u})\bar{p}^A\bar{q}^B
\end{array}\right.\quad (C=0,1,\cdots,n),\end{equation}
where $\bar{u}^C(t,\vartheta)=u^C(t,\theta),\;
\bar{p}^C(t,\vartheta)=p^C(t,\theta)$ and
$\bar{q}^C(t,\vartheta)=q^C(t,\theta)$, respectively.

Introduce the light-cone coordinates
\begin{equation}\label{3.7}\xi=\frac{t+\vartheta}{2},\quad \eta=\frac{t-\vartheta}{2}.\end{equation}
Then, under the coordinates $(\xi,\eta)$, the system (\ref{3.6}) can
be rewritten in the following form
\begin{equation}\label{3.8}\left\{\begin{array}{l}\dfrac{\partial \bar{\bar{p}}^C}{\partial \xi}+
\tilde{\Gamma}^C_{AB}(\bar{\bar{u}})\bar{\bar{p}}^A\bar{\bar{q}}^B=0,\vspace{2mm}\\
\dfrac{\partial \bar{\bar{q}}^C}{\partial
\eta}+\tilde{\Gamma}^C_{AB}(\bar{\bar{u}})\bar{\bar{p}}^A\bar{\bar{q}}^B=0
\end{array}\right.\quad (C=0,1,\cdots,n),\end{equation}
where $\bar{\bar{u}}^C(\xi,\eta)=\bar{u}^C(t,\vartheta),\;
\bar{\bar{p}}^C(\xi,\eta)=\bar{p}^C(t,\vartheta)$ and
$\bar{\bar{q}}^C(\xi,\eta)=\bar{q}^C(t,\vartheta)$, respectively.

On the other hand, noting (\ref{2.8}), (\ref{2.25}), (\ref{3.5}) and
using (\ref{3.7}) gives
\begin{equation}\label{3.9} \bar{\bar{p}}(\xi,\eta)=\bar{\bar{u}}_{\eta}, \quad
\bar{\bar{q}}(\xi,\eta)=\bar{\bar{u}}_{\xi}.\end{equation} Then, the
system (\ref{3.8}) can be equivalently rewritten as
\begin{equation}\label{3.10}
\dfrac{\partial^2 \bar{\bar{u}}^C}{\partial \xi\partial\eta}+
\tilde{\Gamma}^C_{AB}(\bar{\bar{u}})\bar{\bar{u}}_{\eta}^A\bar{\bar{u}}_{\xi}^B=0\quad
(C=0,1,\cdots,n).\end{equation} The equations in (\ref{3.10}) are
nothing but the equations for the wave map from the Minkowski plane
$\mathbb{R}^{1+1}$ to the Lorentzian manifold $(\mathscr{N}, \tilde
g)$. Here we would like to mention that, when the target manifold is
Riemannian instead of Lorentzian, the global existence of smooth
solutions to the system (\ref{3.10}) has been proved by Gu \cite{g2}
successfully.

Recently, Ori \cite{ori} presented a class of curved space-time
vacuum solutions which develop closed timelike curves at some
particular moment, and then used those vacuum solutions to construct
a time-machine model. His solution reads
\begin{equation}\label{3.11}ds^2=dx^2+dy^2-2dzdt+[f(x,y,z)-t]dz^2,\end{equation}
where $f(x,y,z)$ is an arbitrary function (probably periodic in $z$)  satisfying
\begin{equation}\label{3.12}f_{xx}+f_{yy}=0.\end{equation}
In metric (\ref{3.11}), $(t,x,y,z)$ stands for the local coordinates
$(x^0,x^1,x^2,x^3)$ of the enveloping space-time, i.e., the Ori's
space-time in the present situation, and by (\ref{2.8}), also the
coordinates $(u^0,u^1,u^2,u^3)$.

For the case of the Ori's space-time, the system (\ref{3.8}) becomes
\begin{equation}\label{3.13}\left\{\begin{aligned}&\dfrac{\partial \bar{\bar{p}}^0}{\partial \xi}+
\dfrac12(\bar{\bar{p}}^0\bar{\bar{q}}^3+\bar{\bar{p}}^3\bar{\bar{q}}^0)-
\dfrac12f_x(\bar{\bar{p}}^1\bar{\bar{q}}^3+\bar{\bar{p}}^3\bar{\bar{q}}^1)\\&\quad\;\;\;-\dfrac12f_y(\bar{\bar{p}}^2\bar{\bar{q}}^3+\bar{\bar{p}}^3\bar{\bar{q}}^2)+
\dfrac12(t-f-f_z)\bar{\bar{p}}^3\bar{\bar{q}}^3=0,\vspace{2mm}\\
&\dfrac{\partial \bar{\bar{p}}^1}{\partial \xi}-\dfrac12f_x\bar{\bar{p}}^3\bar{\bar{q}}^3=0,\vspace{2mm}\\
&\dfrac{\partial \bar{\bar{p}}^2}{\partial \xi}-\dfrac12f_y\bar{\bar{p}}^3\bar{\bar{q}}^3=0,\vspace{2mm}\\
&\dfrac{\partial \bar{\bar{p}}^3}{\partial
\xi}-\dfrac12\bar{\bar{p}}^3\bar{\bar{q}}^3=0\end{aligned}\right.\end{equation}
and
\begin{equation}\label{3.14}\left\{\begin{aligned}&\dfrac{\partial \bar{\bar{q}}^0}{\partial \eta}+
\dfrac12(\bar{\bar{p}}^0\bar{\bar{q}}^3+\bar{\bar{p}}^3\bar{\bar{q}}^0)-\dfrac12f_x(\bar{\bar{p}}^1\bar{\bar{q}}^3+
\bar{\bar{p}}^3\bar{\bar{q}}^1)\\&\quad\;\;\;-\dfrac12f_y(\bar{\bar{p}}^2\bar{\bar{q}}^3+\bar{\bar{p}}^3\bar{\bar{q}}^2)
+\dfrac12(t-f-f_z)\bar{\bar{p}}^3\bar{\bar{q}}^3=0,\vspace{2mm}\\
&\dfrac{\partial \bar{\bar{q}}^1}{\partial \eta}-\dfrac12f_x\bar{\bar{p}}^3\bar{\bar{q}}^3=0,\vspace{2mm}\\
&\dfrac{\partial \bar{\bar{q}}^2}{\partial \eta}-\dfrac12f_y\bar{\bar{p}}^3\bar{\bar{q}}^3=0,\vspace{2mm}\\
&\dfrac{\partial \bar{\bar{q}}^3}{\partial
\eta}-\dfrac12\bar{\bar{p}}^3\bar{\bar{q}}^3=0,\end{aligned}\right.\end{equation}
while the system (\ref{3.10}) becomes
\begin{equation}\label{3.15}\left\{\begin{aligned}&
\dfrac{\partial^2 \bar{\bar{u}}^0}{\partial \xi\partial\eta}+
\dfrac12(\bar{\bar{u}}_{\eta}^0\bar{\bar{u}}_{\xi}^3+\bar{\bar{u}}_{\eta}^3\bar{\bar{u}}_{\xi}^0)-
\dfrac12f_x(\bar{\bar{u}}_{\eta}^1 \bar{\bar{u}}_{\xi}^3+
\bar{\bar{u}}_{\eta}^3\bar{\bar{u}}_{\xi}^1)\\&\qquad\;\,-\dfrac12f_y(\bar{\bar{u}}_{\eta}^2\bar{\bar{u}}_{\xi}^3+
\bar{\bar{u}}_{\eta}^3\bar{\bar{u}}_{\xi}^2)
+\dfrac12(t-f-f_z)\bar{\bar{u}}_{\eta}^3\bar{\bar{u}}_{\xi}^3=0,\vspace{2mm}\\
&\dfrac{\partial^2 \bar{\bar{u}}^1}{\partial \xi\partial\eta}-
\dfrac12f_x\bar{\bar{u}}_{\eta}^3\bar{\bar{u}}_{\xi}^3=0,\vspace{2mm}\\
&\dfrac{\partial^2 \bar{\bar{u}}^2}{\partial \xi\partial\eta}-
\dfrac12f_y\bar{\bar{u}}_{\eta}^3\bar{\bar{u}}_{\xi}^3=0,\vspace{2mm}\\
&\dfrac{\partial^2 \bar{\bar{u}}^3}{\partial
\xi\partial\eta}-\dfrac12\bar{\bar{u}}_{\eta}^3\bar{\bar{u}}_{\xi}^3=0.\end{aligned}\right.\end{equation}

As in Ori \cite{ori}, for concreteness we now specialize to a simple
example. Take
\begin{equation}\label{3.16}
f=a(x^2-y^2)
\end{equation}
for some positive constant $a$. This yields an empty curved
space-time, locally isometric to a linearly polarized plane wave.
For this concrete situation, the system (\ref{3.15}) becomes
\begin{equation}\label{3.17}\left\{\begin{aligned}&
\dfrac{\partial^2 \bar{\bar{u}}^0}{\partial \xi\partial\eta}+
\dfrac12(\bar{\bar{u}}_{\eta}^0\bar{\bar{u}}_{\xi}^3+\bar{\bar{u}}_{\eta}^3\bar{\bar{u}}_{\xi}^0)-
a\bar{\bar{u}}^1(\bar{\bar{u}}_{\eta}^1 \bar{\bar{u}}_{\xi}^3+
\bar{\bar{u}}_{\eta}^3\bar{\bar{u}}_{\xi}^1)\\&\qquad\;\,+a\bar{\bar{u}}^2
(\bar{\bar{u}}_{\eta}^2\bar{\bar{u}}_{\xi}^3+
\bar{\bar{u}}_{\eta}^3\bar{\bar{u}}_{\xi}^2)
+\dfrac12\left\{\bar{\bar{u}}^0-a\left[(\bar{\bar{u}}^1)^2-(\bar{\bar{u}}^2)^2\right]\right\}
\bar{\bar{u}}_{\eta}^3\bar{\bar{u}}_{\xi}^3=0,\vspace{2mm}\\
&\dfrac{\partial^2 \bar{\bar{u}}^1}{\partial \xi\partial\eta}-
a\bar{\bar{u}}^1\bar{\bar{u}}_{\eta}^3\bar{\bar{u}}_{\xi}^3=0,\vspace{2mm}\\
&\dfrac{\partial^2 \bar{\bar{u}}^2}{\partial \xi\partial\eta}+a\bar{\bar{u}}^2
\bar{\bar{u}}_{\eta}^3\bar{\bar{u}}_{\xi}^3=0,\vspace{2mm}\\
&\dfrac{\partial^2 \bar{\bar{u}}^3}{\partial
\xi\partial\eta}-\dfrac12\bar{\bar{u}}_{\eta}^3\bar{\bar{u}}_{\xi}^3=0.\end{aligned}\right.\end{equation}
The key point to solve the system (\ref{3.17}) is to solve
$\bar{\bar{u}}^3$ from the last equation in (\ref{3.17}). In fact,
once one solves $\bar{\bar{u}}^3$ from the last equation, then the
second and third equations in (\ref{3.17}) become linear, and then
one can easily solve the unknown functions $\bar{\bar{u}}^1$ and
$\bar{\bar{u}}^2$ from the second and third equations in
(\ref{3.17}). After solving $\bar{\bar{u}}^1$, $\bar{\bar{u}}^2$ and
$\bar{\bar{u}}^3$, by substituting them into the first equation in
(\ref{3.17}), one can find that the first equation in (\ref{3.17})
also becomes linear, and then one can easily solve $\bar{\bar{u}}^0$
from it. Therefore, in what follows, it suffices to consider the
last equation in (\ref{3.17}).

In order to solve the last equation in (\ref{3.17}), denoted by
(\ref{3.17})$_4$, we first gives the corresponding initial data.

In fact, the half plane $\{(t,\vartheta)\,|\,t\ge 0,\;
\vartheta\in\mathbb{R} \}$ in the coordinates $(t,\vartheta)$
becomes the half plane $\{(\xi,\eta)\,|\,\xi+\eta\ge 0 \}$ in the
coordinates $(\xi,\eta)$, while the initial line $t=0$ becomes the
line $\xi+\eta=0$. See Figure 1.

\begin{figure}[!ht]
\centering
\includegraphics{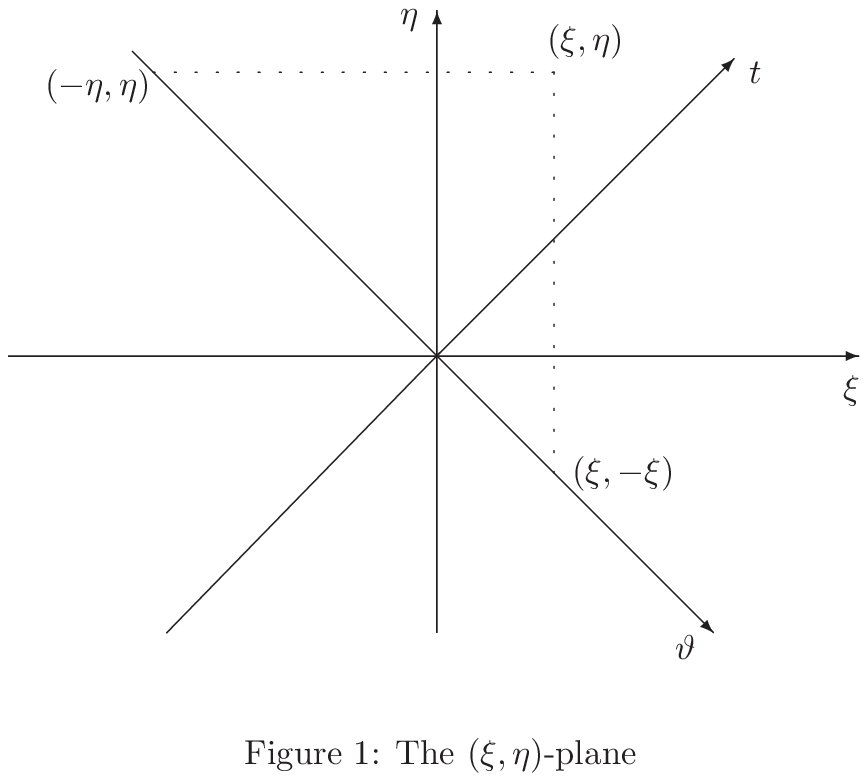}\end{figure}

\noindent Therefore, by (\ref{3.7}), it follows from the initial
data (\ref{2.28}) defined on $t=0$ that
\begin{equation}\label{3.18}
\bar{\bar{u}}^3(\xi,-\xi)=\bar{\varphi}^3(2\xi),\quad
\bar{\bar{u}}^3_{\eta}(\xi,-\xi)=\bar{p}_0^3(2\xi)
\end{equation}
on the initial line $\xi+\eta=0$.

We next solve the Cauchy problem (\ref{3.17})$_4$, (\ref{3.18}) on
the half plane $\{(\xi,\eta)\,|\,\xi+\eta\ge 0 \}$.

Notice that (3.17)$_4$ can be rewritten as
\begin{equation}\label{3.19}
\frac{\partial
\bar{\bar{u}}_{\eta}^3}{\partial\xi}=\dfrac12\bar{\bar{u}}_{\xi}^3\bar{\bar{u}}_{\eta}^3.
\end{equation}
Fixing $\eta$ and integrating (\ref{3.19}) with respect to $\xi$
from $-\eta$ gives
\begin{equation}\label{3.20}\begin{array}{lll}{\displaystyle
\frac{\partial\bar{\bar{u}}^3}{\partial\eta}(\xi,\eta)} & = &
{\displaystyle
\frac{\partial\bar{\bar{u}}^3}{\partial\eta}(-\eta,\eta)\times
\exp\left\{\frac12\int^{\xi}_{-\eta}\frac{\partial\bar{\bar{u}}^3}{\partial
s}(s,\eta)ds\right\}}\vspace{2mm}\\
& = & {\displaystyle \bar{p}_0^3(-2\eta)\times
\exp\left\{\frac12\left(\bar{\bar{u}}^3(\xi,\eta)-\bar{\bar{u}}^3(-\eta,\eta)\right)\right\}}\vspace{2mm}\\
& = & {\displaystyle \bar{p}_0^3(-2\eta)\times
\exp\left\{\frac12\bar{\bar{u}}^3(\xi,\eta)\right\}\times
\exp\left\{-\frac12\bar{\varphi}^3(-2\eta)\right\}.}\end{array}
\end{equation}
Obviously, (3.20) can be rewritten as
\begin{equation}\label{3.21}
\frac{\partial
}{\partial\eta}\left(\exp\left\{-\frac12\bar{\bar{u}}^3(\xi,\eta)\right\}\right)=-\frac12
\bar{p}_0^3(-2\eta)\times
\exp\left\{-\frac12\bar{\varphi}^3(-2\eta)\right\}.
\end{equation}
Fixing $\xi$ and integrating (\ref{3.21}) with respect to $\eta$
from $-\xi$ leads to
\begin{equation}\label{3.22}
\exp\left\{-\frac12\bar{\bar{u}}^3(\xi,\eta)\right\}=\exp\left\{-\frac12\bar{\varphi}^3(2\xi)\right\}
-\frac12\int^{\eta}_{-\xi}\bar{p}^3_0(-2s)\exp\left\{-\frac12\bar{\varphi}^3(-2s)\right\}ds,
\end{equation}
namely,
\begin{equation}\label{3.23}
\bar{\bar{u}}^3(\xi,\eta)=-2\ln\left\{\exp\left\{-\frac12\bar{\varphi}^3(2\xi)\right\}
-\frac12\int^{\eta}_{-\xi}\bar{p}^3_0(-2s)\exp\left\{-\frac12\bar{\varphi}^3(-2s)\right\}ds\right\}.
\end{equation}
In the coordinates $(t,\vartheta)$, (\ref{3.23}) becomes
\begin{equation}\label{3.24}
\bar{{u}}^3(t,\vartheta)=-2\ln\left\{\exp\left\{-\frac12\bar{\varphi}^3(t+\vartheta)\right\}
-\frac12\int^{\frac{t-\vartheta}{2}}_{-\frac{t+\vartheta}{2}}\bar{p}^3_0(-2s)
\exp\left\{-\frac12\bar{\varphi}^3(-2s)\right\}ds\right\}.
\end{equation}
Noting that
\begin{equation}\label{3.25}
\bar{p}^3_0=\bar{\psi}^3-\bar{\varphi}^3_{\vartheta},
\end{equation}
we obtain from (\ref{3.24}) that
\begin{equation}\begin{aligned}\label{3.26}
\bar{{u}}^3(t,\vartheta)=&-2\ln\left\{\frac12\exp\left\{-\frac12\bar{\varphi}^3(t+\vartheta)\right\}
+\frac12\exp\left\{-\frac12\bar{\varphi}^3(\vartheta-t)\right\}\right.\\
&\left.-\frac12\int^{\frac{\vartheta+t}{2}}_{\frac{\vartheta
-t}{2}}\bar{\psi}^3(2s)
\exp\left\{-\frac12\bar{\varphi}^3(2s)\right\}ds\right\}.
\end{aligned}
\end{equation}

Summarizing the above arguments yields

\begin{Theorem} Under the assumptions (\ref{2.33}) and
(\ref{2.34}) (i.e., the physical motion assumptions), the Cauchy
problem for the relativistic string equations in the Ori's
space-time with the initial data (\ref{2.28}) admits a unique global
$C^2$ smooth solution on $\mathbb{R}^+\times \mathbb{R}$ if and only
if, in the coordinates $(t,\vartheta)$ the initial data satisfies
\begin{equation}\label{3.28}
\exp\left\{-\frac12\bar{\varphi}^3(\vartheta+t)\right\}+\exp\left\{-\frac12\bar{\varphi}^3(\vartheta-t)\right\}
>\int^{\frac{\vartheta +t}{2}}_{\frac{\vartheta-t}{2}}\bar{\psi}^3(2s)
\exp\left\{-\frac12\bar{\varphi}^3(2s)\right\}ds,\quad \forall\;
t>0,\;\forall\; \vartheta\in \mathbb{R},
\end{equation}
where $\bar{\varphi}^3$ and $\bar{\psi}^3$ are the corresponding
initial data in the coordinates $(t,\vartheta)$ of $\varphi^3$ and
$\psi^3$ in the original coordinates $(t,\theta)$, respectively.
\end{Theorem}

\begin{Corollary}
If $\psi^3(\theta)\le 0$ for all $\theta\in \mathbb{R}$, then the
Cauchy problem for the relativistic string equations in the Ori's
space-time with the initial data (\ref{2.28}) admits a unique global
$C^2$ smooth solution on $\mathbb{R}^+\times \mathbb{R}$.
\end{Corollary}

On the other hand, noting (\ref{3.24}), we have
\begin{Corollary}
If $p^3_0(\theta)\le 0$ for all $\theta\in \mathbb{R}$, then the
Cauchy problem for the relativistic string equations in the Ori's
space-time with the initial data (\ref{2.28}) admits a unique global
$C^2$ smooth solution on $\mathbb{R}^+\times \mathbb{R}$.
\end{Corollary}

\begin{Remark} It is obvious that (3.17)$_4$ can also be rewritten as
\begin{equation}\label{3.27}
\frac{\partial
\bar{\bar{u}}_{\xi}^3}{\partial\eta}=\dfrac12\bar{\bar{u}}_{\eta}^3\bar{\bar{u}}_{\xi}^3,
\end{equation}
in this case we have a similar discussion by means of $\bar{q}_0^3$
instead of $\bar{p}_0^3$. That is to say, in this way we can also
prove Theorem 3.1.
\end{Remark}

Thus, using (\ref{3.24}) and a similar one by means of
$\bar{q}_0^3$, we have
\begin{Corollary}
If it holds that
$$\|\bar{p}^3_0\|_{L^1}\ll 1\quad and \quad \|\bar{q}^3_0\|_{L^1}\ll 1
,$$ then the Cauchy problem for the relativistic string equations in
the Ori's space-time with the initial data (\ref{2.28}) admits a
unique global $C^2$ smooth solution on $\mathbb{R}^+\times
\mathbb{R}$.
\end{Corollary}

\section{Conclusion and discussion}
It is well known, in particle physics, the string model is used to
consider the structure of hardrons. A free string is a
one-dimensional physical object whose motion is represented by a
time-like extremal surface in the physical space-times. The extremal
surfaces play an important role in both mathematics and physics, in
particular, in the theoretical apparatus of elementary particle
physics. The theory on the motion of a relativistic string in the
Minkowski spacetime has been studied extensively, many and beautiful
results have been obtained. However, the study on the motion of a
relativistic string in curved space-times is vastly open, there are
a lot of fundamentally important problems needed to solve. The main
difficulty is that the PDEs for such a motion are essentially
nonlinear.

In fact, the motion of a string in a curved enveloping space-time
$(\mathscr{N}, \tilde g)$, which stands for a given general
Lorentzian manifold, can be determined by constructing a certain
wave map from the Minkowski plane to $(\mathscr{N}, \tilde g)$. When
the target manifold is Riemannian, the global existence of smooth
solutions to the system (\ref{3.10}) has been proved by Gu
\cite{g2}. On the other hand, when the enveloping space-time
$(\mathscr{N}, \tilde g)$ takes some special cases, for example,
$\mathbb{S}^{1+1}$ or the Schwarzchild space-time, some results on
the global existence of smooth solutions to the corresponding wave
map equations have also been obtained (see \cite{g10}, \cite{hek}).
In the present paper, we consider another special but important case
that the enveloping space-time is Ori's: (1) as a general framework,
we first analyze relativistic string equations in a curved
enveloping space-time $(\mathscr{N}, \tilde g)$ which stands for a
general Lorentzian manifold, and then investigate some interesting
properties enjoyed by these equations; (2) based on this, under
suitable {\it small} assumptions we prove the global existence of
smooth solutions of the Cauchy problem for relativistic string
equations in  $(\mathscr{N}, \tilde g)$; (3) in particular, we
investigate the motion of a relativistic string in the Ori's
space-time, and give a sufficient and necessary condition
guaranteeing the global existence of smooth solutions of the Cauchy
problem for relativistic string equations in the Ori's space-time.

Our ultimate goal is to study the global existence or breakdown
phenomena of smooth solutions of the relativistic string equations
in a general curved enveloping space-time $(\mathscr{N}, \tilde g)$
without any {\it small} assumption. This is a hard task. However, in
the Gaussian coordinates, the Lorentzian metric of the (curved)
space-time described by the Einstein's field equations can be
written, at least locally, as
\begin{equation}\label{4.1}\tilde
g=\left( \begin{array}{cc} -1 & 0 \\ 0 & h
\end{array}\right),\end{equation}
where $=(h_{ij})_{n\times n}$ stands for a Riemannian metric. See
Kossowski and Kriele \cite{kk} for the details. In the present
situation, the system (\ref{3.6}) becomes
 \begin{equation}\label{4.2}\left\{\begin{array}{l}{\displaystyle\dfrac{\partial p^0}{\partial t}+
 \dfrac{\partial p^0}{\partial \vartheta}=-
\frac12\frac{\partial h_{ij}}{\partial x^0}p^iq^j,}\vspace{3mm}\\
{\displaystyle\dfrac{\partial p^k}{\partial t}+\dfrac{\partial
p^k}{\partial \vartheta}=-\frac12 h^{kl}\frac{\partial
h_{li}}{\partial x^0}p^0q^i-\frac12h^{kl}\frac{\partial
h_{li}}{\partial x^0}p^iq^0-\hat{\Gamma}^k_{ij}p^iq^j}\quad
(k=1,\cdots,n),
\end{array}\right.\end{equation}
where $\hat{\Gamma}^k_{ij}$ stand for the connections corresponding
to the Riemannian metric $h$. Similarly, the equations satisfied by
$q$ can be obtained. It is easy to see tat the system (\ref{4.2})
somewhat possesses a special form with some geometric structures,
perhaps this will shed light on solving our ultimate problem. This
is worthy to be studied seriously in the future.

\vskip 10mm

\noindent{\Large {\bf Acknowledgements.}} This work was completed
while Kong was visiting the Max Planck Institute for Gravitational
Physics (Albert Einstein Institute) during the summer of 2010. Kong
thanks L. Andersson for his invitation and hospitality. This work
was supported in part by the NNSF of China (Grant No. 10971190), the
Qiu-Shi Chair Professor Fellowship from Zhejiang University, the
Foundation for University's Excellent Youth Scholars (Grant Nos.
2009SQRZ025ZD, 2010SQRL025) and the University's Natural Science
Foundation from Anhui Province (Grand No. KJ2010A130).

\end{document}